\documentclass{PoS}

\usepackage{textpos}
\usepackage{feynmp-auto}
\usepackage{subcaption}
\usepackage[squaren,mediumqspace]{SIunits}

\newcommand{\amuSUSY}{a_{\mu}^{\text{SUSY}}}
\newcommand{\amured}{a_{\mu}^{\text{red}}}
\newcommand{\Deltamured}{\Delta_\mu^{\text{red}}}

\def\lsim{\mathrel{\rlap{\lower4pt\hbox{$\sim$}}
    \raise2pt\hbox{$<$}}}                
\def\gsim{\mathrel{\rlap{\lower4pt\hbox{$\sim$}}
    \raise2pt\hbox{$>$}}}                
\newcommand{\Mone}{M_{1}}
\newcommand{\Mtwo}{M_{2}}
\newcommand{\ML}{m_{L}}
\newcommand{\MR}{m_{R}}
\newcommand{\MUE}{\mu}
\newcommand{\GeV}{\ \mathrm{GeV}}

\newcommand{\TeV}{\ \mathrm{TeV}}
\newcommand{\tb}{\tan\!\beta}
\newcommand{\amuSUSYOL}{a_{\mu}^{\mathrm{SUSY,1L}}}
\newcommand{\Deltared}[1]{\Delta_{#1}^{\mathrm{red}}}
\newcommand{\MSUSY}{M_{\mathrm{SUSY}}}
\newcommand{\MSUSYmin}{M_{\mathrm{SUSY,min}}}
\newcommand{\sign}{\mathop{\mathrm{sign}}}

\newcommand{\graphwhn}[6]{%
  \begin{fmffile}{#1}
    \begin{fmfgraph*}(100,100)
      \fmfleft{muL}
      \fmfright{muR}
      \fmffixed{(.10w,.25h)}{v1,x1}
      \fmffixed{(-.10w,.25h)}{v2,x3}
      \fmffixed{(.35w,.35h)}{v1,x2}
      \fmffixed{(-.35w,.35h)}{v2,x2}
      \fmf{plain,label=$\mu_L$}{muL,v1}
      \fmf{plain,label=$\mu_R$}{v2,muR}
      \fmf{dashes,tension=0.3,label=$\widetilde{\nu}_\mu$,f=#5}{v1,v2}
      \fmf{plain,left=.2,tension=0,label=$\widetilde{W}^-$,f=#4}{v1,x1}
      \fmf{plain,left=.2,tension=0,label=$\widetilde{W}^-$,f=#4}{x1,x2}
      \fmf{plain,left=.2,tension=0,label=$\widetilde{H}_u^-$,f=#2}{x2,x3}
      \fmf{plain,left=.2,tension=0,label=$\widetilde{H}_d^-$,f=#2}{x3,v2}
      \fmfv{dec.shape=cross,dec.size=10,f=blue}{x1}
      \fmfv{dec.shape=cross,dec.size=10,f=blue}{x2}
      \fmfv{dec.shape=cross,dec.size=10,f=blue}{x3}
    \end{fmfgraph*}
  \end{fmffile}%
}
\newcommand{\graphwhmL}[6]{%
  \begin{fmffile}{#1}
    \begin{fmfgraph*}(100,100)
      \fmfleft{muL}
      \fmfright{muR}
      \fmffixed{(.10w,.25h)}{v1,x1}
      \fmffixed{(-.10w,.25h)}{v2,x3}
      \fmffixed{(.35w,.35h)}{v1,x2}
      \fmffixed{(-.35w,.35h)}{v2,x2}
      \fmf{plain,label=$\mu_L$}{muL,v1}
      \fmf{plain,label=$\mu_R$}{v2,muR}
      \fmf{dashes,tension=0.3,label=$\widetilde{\mu}_L$,f=#5}{v1,v2}
      \fmf{plain,left=.2,tension=0,label=$\widetilde{W}^0$,f=#4}{v1,x1}
      \fmf{plain,left=.2,tension=0,label=$\widetilde{W}^0$,f=#4}{x1,x2}
      \fmf{plain,left=.2,tension=0,label=$\widetilde{H}_u^0$,f=#2}{x2,x3}
      \fmf{plain,left=.2,tension=0,label=$\widetilde{H}_d^0$,f=#2}{x3,v2}
      \fmfv{dec.shape=cross,dec.size=10,f=blue}{x1}
      \fmfv{dec.shape=cross,dec.size=10,f=blue}{x2}
      \fmfv{dec.shape=cross,dec.size=10,f=blue}{x3}
    \end{fmfgraph*}
  \end{fmffile}%
}
\newcommand{\graphbhmL}[6]{%
  \begin{fmffile}{#1}
    \begin{fmfgraph*}(100,100)
      \fmfleft{muL}
      \fmfright{muR}
      \fmffixed{(.10w,.25h)}{v1,x1}
      \fmffixed{(-.10w,.25h)}{v2,x3}
      \fmffixed{(.35w,.35h)}{v1,x2}
      \fmffixed{(-.35w,.35h)}{v2,x2}
      \fmf{plain,label=$\mu_L$}{muL,v1}
      \fmf{plain,label=$\mu_R$}{v2,muR}
      \fmf{dashes,tension=0.3,label=$\widetilde{\mu}_L$,f=#5}{v1,v2}
      \fmf{plain,left=.2,tension=0,label=$\widetilde{B}$,f=#3}{v1,x1}
      \fmf{plain,left=.2,tension=0,label=$\widetilde{B}$,f=#3}{x1,x2}
      \fmf{plain,left=.2,tension=0,label=$\widetilde{H}_u^0$,f=#2}{x2,x3}
      \fmf{plain,left=.2,tension=0,label=$\widetilde{H}_d^0$,f=#2}{x3,v2}
      \fmfv{dec.shape=cross,dec.size=10,f=blue}{x1}
      \fmfv{dec.shape=cross,dec.size=10,f=blue}{x2}
      \fmfv{dec.shape=cross,dec.size=10,f=blue}{x3}
    \end{fmfgraph*}
  \end{fmffile}%
}
\newcommand{\graphbhmR}[6]{%
  \begin{fmffile}{#1}
    \begin{fmfgraph*}(100,100)
      \fmfleft{muL}
      \fmfright{muR}
      \fmffixed{(.10w,.25h)}{v1,x1}
      \fmffixed{(-.10w,.25h)}{v2,x3}
      \fmffixed{(.35w,.35h)}{v1,x2}
      \fmffixed{(-.35w,.35h)}{v2,x2}
      \fmf{plain,label=$\mu_L$}{muL,v1}
      \fmf{plain,label=$\mu_R$}{v2,muR}
      \fmf{dashes,tension=0.3,label=$\widetilde{\mu}_R$,f=#6}{v1,v2}
      \fmf{plain,left=.2,tension=0,label=$\widetilde{H}_d^0$,f=#2}{v1,x1}
      \fmf{plain,left=.2,tension=0,label=$\widetilde{H}_u^0$,f=#2}{x1,x2}
      \fmf{plain,left=.2,tension=0,label=$\widetilde{B}$,f=#3}{x2,x3}
      \fmf{plain,left=.2,tension=0,label=$\widetilde{B}$,f=#3}{x3,v2}
      \fmfv{dec.shape=cross,dec.size=10,f=blue}{x1}
      \fmfv{dec.shape=cross,dec.size=10,f=blue}{x2}
      \fmfv{dec.shape=cross,dec.size=10,f=blue}{x3}
    \end{fmfgraph*}
  \end{fmffile}%
}
\newcommand{\graphbmLmR}[6]{%
  \begin{fmffile}{#1}
    \begin{fmfgraph*}(100,100)
      \fmfleft{muL}
      \fmfright{muR}
      \fmffixed{(.10w,.25h)}{v1,x1}
      \fmffixed{(-.10w,.25h)}{v2,x3}
      \fmffixed{(.35w,.35h)}{v1,x2}
      \fmffixed{(-.35w,.35h)}{v2,x2}
      \fmf{plain,label=$\mu_L$}{muL,v1}
      \fmf{plain,label=$\mu_R$}{v2,muR}
      \fmf{dashes,tension=0.3,label=$\widetilde{\mu}_L$,f=#5}{v1,x4}
      \fmf{dashes,tension=0.3,label=$\widetilde{\mu}_R$,label.side=right,f=#6}{x4,v2}
      \fmf{plain,left=.4,tension=0,label=$\widetilde{B}$,f=#3}{v1,x2}
      \fmf{plain,left=.4,tension=0,label=$\widetilde{B}$,f=#3}{x2,v2}
      \fmfv{dec.shape=cross,dec.size=10,f=blue}{x2}
      \fmfv{dec.shape=cross,dec.size=10,f=blue}{x4}
    \end{fmfgraph*}
  \end{fmffile}%
}

\newcommand{\graphsAsSubfigs}[6]{%
  \begin{subfigure}{.3\textwidth}
    \centering
    \graphwhn{graph-whn-#1}{#2}{#3}{#4}{#5}{#6}\vspace{-7ex}
    \caption{}
  \end{subfigure}
  \begin{subfigure}{.3\textwidth}
    \centering
    \graphwhmL{graph-whmL-#1}{#2}{#3}{#4}{#5}{#6}\vspace{-7ex}
    \caption{}
  \end{subfigure}
  \\[2ex]
  \begin{subfigure}{.3\textwidth}
    \centering
    \graphbhmL{graph-bhmL-#1}{#2}{#3}{#4}{#5}{#6}\vspace{-7ex}
    \caption{}
  \end{subfigure}
  \begin{subfigure}{.3\textwidth}
    \centering
    \graphbhmR{graph-bhmR-#1}{#2}{#3}{#4}{#5}{#6}\vspace{-7ex}
    \caption{}
  \end{subfigure}
  \begin{subfigure}{.3\textwidth}
    \centering
    \graphbmLmR{graph-bmLmR-#1}{#2}{#3}{#4}{#5}{#6}\vspace{-7ex}
    \caption{}
  \end{subfigure}
}

\title{\boldmath
 Large muon $(g-2)$ from TeV-scale MSSM with infinite $\tb$}

\ShortTitle{Large muon $(g-2)$ from TeV-scale MSSM with infinite $\tb$}

\author{\speaker{Jae-hyeon Park}\\ 
        Departament de F\'{i}sica Te\`{o}rica and IFIC,
Universitat de Val\`{e}ncia-CSIC,
46100, Burjassot, Spain\\
        E-mail: \email{jae.park@uv.es}}

\author{Markus Bach, Dominik St\"ockinger and Hyejung St\"ockinger-Kim\\
        Institut f\"ur Kern- und Teilchenphysik,
TU Dresden, 01069 Dresden, Germany\\
        E-mail: \email{markus.bach1@tu-dresden.de},
                \email{dominik.stoeckinger@tu-dresden.de},
                \email{hyejung.stoeckinger-kim@tu-dresden.de}}

\abstract{%
  \begin{textblock*}{7em}(0.82\textwidth,-0.63\textheight)
    \noindent\footnotesize
    FTUV--15--1070 \\
    IFIC--15--75 \\
  \end{textblock*}%
  The muon anomalous magnetic moment $a_\mu$ is studied
  in the infinite $\tb$ limit of the MSSM\@.
  Since the muon mass arises completely from loop effects,
  large corrections to $a_\mu$ are expected in comparison to
  the usual case with moderately high $\tb$.
  Due to the qualitatively different parameter dependence,
  the gap between the experimental value and the Standard
  Model prediction can be filled only in parameter volumes in which
  a mass hierarchy suppresses the chargino loop
  in favour of the neutralino contribution.
  Two such possibilities are found to have either large Higgsino mass or
  large muon sneutrino mass.
  Supersymmetric particles even at the TeV scale can lead to
  the best fit of $a_\mu$.}

\FullConference{18th International Conference From the Planck Scale to the Electroweak Scale \\
		25-29 May 2015\\
		Ioannina, Greece }

\graphicspath{{./}{../}}

\begin{document}

\noindent
The muon anomalous magnetic moment, $a_\mu\equiv(g_\mu-2)/2$,
has long been a prime observable
sensitive to hypothetical virtual states, and hence
a major indirect probe to new physics.
Currently, there is an interesting discrepancy between the
experimental and the Standard Model (SM) values of $a_\mu$ \cite{Davier}:
\begin{equation}
  a_\mu^{\mathrm{exp}} - a_\mu^{\mathrm{SM}} =
  (28.7 \pm 8.0) \times 10^{-10} ,
\label{eq:amuexp-amusm}
\end{equation}
which amounts to more than $3\sigma$.
This has called for many attempts to fill the gap
with various new physics contributions.

\begin{figure}
  \centering
    \begin{fmffile}{graph-gm2}
      \begin{fmfgraph*}(100,80)
        \fmfleft{i1}
        \fmfright{o1}
        \fmfbottom{p1}
        \fmf{plain,label=$\mu_L$}{i1,v}
        \fmf{plain,label=$\mu_R$}{v,o1}
        \fmfv{d.sh=circle,d.f=empty,d.si=.4w,b=(.8,,.8,,.8)}{v}
        \fmffreeze
        \fmf{photon}{v,p1}
      \end{fmfgraph*}
    \end{fmffile}
    \quad \raisebox{7.5ex}{$\longleftrightarrow$}\quad
    \begin{fmffile}{graph-self}
      \begin{fmfgraph*}(100,80)
        \fmfleft{i1}
        \fmfright{o1}
        \fmf{plain,label=$\mu_L$}{i1,v}
        \fmf{plain,label=$\mu_R$}{v,o1}
        \fmfv{d.sh=circle,d.f=empty,d.si=.4w,b=(.8,,.8,,.8),l=$\Sigma_\mu$,l.dist=-4}{v}
      \end{fmfgraph*}
    \end{fmffile}
  \caption{Diagrammatic similarity between $a_\mu$ (left)
    and the muon self energy (right).}
  \label{fig:amu and muon self energy}
\end{figure}
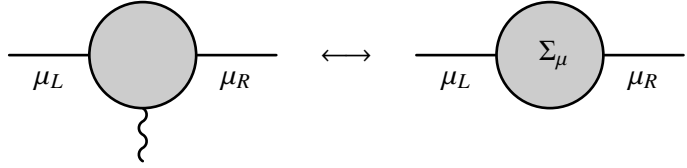

In this respect,
one noteworthy property of $a_\mu$ is that
it is correlated with loop corrections to the muon mass $m_\mu$.
This is easy to see by comparing the schematic diagrams of
$a_\mu$ and the muon self energy $\Sigma_\mu$,
shown in figure~\ref{fig:amu and muon self energy}.
Manifestly, attaching a photon line to
the $\Sigma_\mu$ graph on the right
results in the $a_\mu$ diagram on the left.
This means that large loop corrections to $m_\mu$ generically imply
large contributions to $a_\mu$.

For this reason, there have been significant interests in
models where $m_\mu$ arises solely from loop effects.
This idea may be realized for instance
within a popular model such as
the minimal supersymmetric standard model (MSSM)\@.
In this model,
the tree-level muon mass is given by
\begin{equation}
  m_\mu^\mathrm{tree} = y_\mu v_d ,
\end{equation}
the product of $y_\mu$, the muon Yukawa coupling, and
$v_d$, the vacuum expectation value (VEV) of the down-type Higgs.
Obviously, this leads to the following two options
to eliminate the tree-level muon mass:
(a) $y_\mu = 0$ or (b) $v_d = 0$.
There are already studies which employ
the former approach while using
the non-holomorphic \cite{nonholo}
or the holomorphic \cite{Thalapillil:2014kya}
smuon trilinear coupling
for the radiative generation of $m_\mu$.
We consider the latter option
focusing on the supersymmetric contributions to $a_\mu$ \cite{Bach:2015doa}
which shall be the subject of this presentation.


The supersymmetric contributions to $a_\mu$ at the one-loop level
are well-known in the MSSM \cite{moroi}.
They arise from the chargino-sneutrino and the
neutralino-smuon diagrams which depend on the following five mass parameters:
the Higgsino mass $\mu$,
the bino mass $\Mone$,
the wino mass $\Mtwo$ as well as
the soft masses of the left- and the right-handed smuons, $m_L$ and $m_R$.
In a simplified case where all these five parameters are equal to
$\MSUSY$,
the above contributions can be approximated by
\begin{equation}
  a_\mu^\mathrm{SUSY,1L} \approx
  13 \times
  10^{-10} \,\sign(\mu) 
  \tb \left(\frac{\unit{100}{\giga\electronvolt}}{\MSUSY}\right)^2 .
\label{eq:amuSUSY1L}
\end{equation}
One can notice the following properties of $\amuSUSYOL$:
(a) it is suppressed by the second power of $\MSUSY$, the new physics scale,
(b) it is proportional to $\tb$,
(c) its sign is determined by the sign of $\mu$.
For a usual value of $\tb \lesssim 60$,
the best fit of (\ref{eq:amuexp-amusm}) is achieved only if
$\MSUSY \lesssim 500\GeV$.
Now that the Large Hadron Collider is putting more and more
stringent lower bounds on the supersymmetric particle masses,
one may consider how to fit (\ref{eq:amuexp-amusm})
even for higher $\MSUSY$.

To this end, a straightforward attempt would be to push $\tb$ up
beyond the usual range.
This has been however a less explored possibility.
The fear is mostly based on
the perturbativity of the down-type Yukawa couplings.
As they are proportional to $\tb$ under the tree-level approximation,
one might naively expect them to grow nonperturbatively large
for too high $\tb$.
It is however well-known that
the down-type fermion masses can receive significant loop corrections
which pick up the up-type Higgs VEV $v_u$ \cite{Hall:1993gn}.
This effect may be expressed in the form,
\begin{equation}
  m_f = y_f\, v_d + y_f\, v_u \,\Deltared{f} ,
  \label{eq:mass including loops}
\end{equation}
where $m_f$ is the pole mass of $f$
standing for a charged lepton or down-type quark,
$y_f$ is its Yukawa coupling,
and $\Deltared{f}$ is a finite quantity calculable from
the self energy diagrams.
This implies that $m_f$ might be correct even if
$v_d$ is unconventionally small or even vanishes.
Indeed, it has been shown that such a possibility can
meet various phenomenological constraints
\cite{Dobrescu:2010mk,Altmannshofer:2010zt}.

With $\Deltared{f}$ from (\ref{eq:mass including loops}) taken
into account,
the supersymmetric contribution to $a_\mu$ becomes
\cite{Marchetti:2008hw}
\begin{equation}
  \amuSUSY =
             \frac{\amuSUSYOL}{1 + \tb\Deltamured}
             \equiv \frac{y_\mu v_u}{m_\mu} \amured ,
  \label{eq:resummed amuSUSY}
\end{equation}
where $\amuSUSYOL$, appearing on the left-hand side of (\ref{eq:amuSUSY1L}),
is the supersymmetric one-loop contribution to $a_\mu$
without resummation of $\tb$-enhanced terms.
One can understand the first equality above in the following way.
Due to the chiral symmetry,
both $\amuSUSY$ and $\amuSUSYOL$
contain a factor of $y_\mu$ which is rewritten in terms of the other
quantities appearing in (\ref{eq:mass including loops}).
This replacement is done however to differing accuracies, i.e.\
the $\Deltared{f}$ term in (\ref{eq:mass including loops})
is neglected for $\amuSUSYOL$ whereas it is included for $\amuSUSY$.
The above resummation formula then follows.
Another quantity $\amured$ is defined in (\ref{eq:resummed amuSUSY}),
in terms of which the infinite $\tb$ limit of $\amuSUSY$
reads simply
\begin{equation}
  \lim_{\tan\beta\to\infty} \amuSUSY = \frac{\amured}{\Deltamured} ,
  \label{eq:amuSUSYratio}
\end{equation}
which follows from (\ref{eq:mass including loops}) and
(\ref{eq:resummed amuSUSY}).


In this limit,
one can again consider a simplified case
where the five mass parameters are the same
to obtain an expression analogous to (\ref{eq:amuSUSY1L}).
This results in
\begin{equation}
  \lim_{\tb\to\infty} \amuSUSY \approx
  {\mathbf{-}} 72 \times 10^{-10} \,
  \left(\frac{\unit{1}{\tera\electronvolt}}{\MSUSY}\right)^2 .
  \label{eq:negativeamuSUSY}
\end{equation}
For the same new physics scale $\MSUSY$,
the magnitude of $\amuSUSY$ above does greatly exceed
that of (\ref{eq:amuSUSY1L}) valid for moderate $\tb$.
Unfortunately, the sign turns out to be wrong
which, being independent of the $\mu$ sign, seems unfixable.
This is because the diagrams of both $\amured$ and $\Deltamured$
pick up a Higgsino mass due to the Peccei-Quinn symmetry
and therefore the $\mu$ sign cancels out of (\ref{eq:amuSUSYratio}).

\begin{figure}
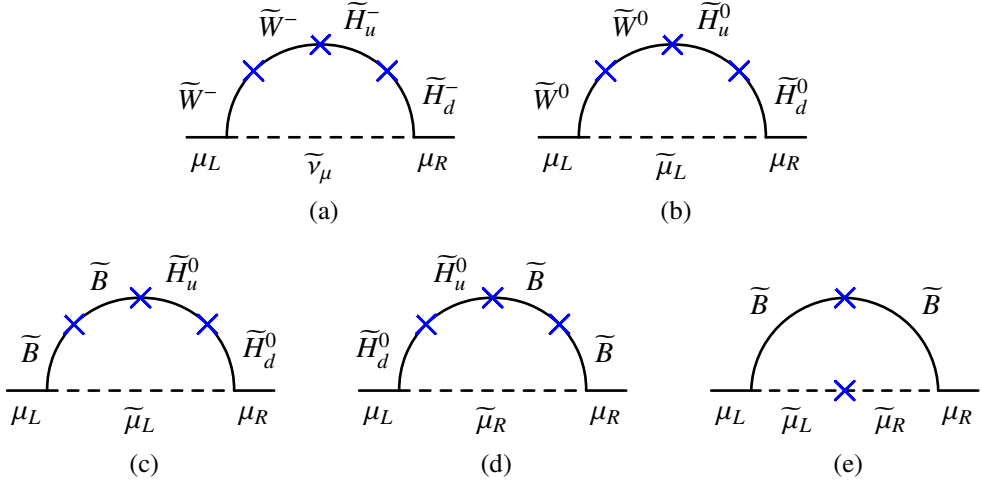

  \centering
  \graphsAsSubfigs{normal}{black}{black}{black}{black}{black}
  \caption{Mass-insertion diagrams for $\Deltamured$ and $\amured$.
    For the latter, an external photon couples to any of the
    charged particles in the loop.}
  \label{fig:diagrams}
\end{figure}

It is still early to give up.
One can get a hint on how to overcome this sign problem
by closer inspection of the structures of $\amured$ and $\Deltamured$
appearing in the fraction of~(\ref{eq:amuSUSYratio}).
Schematically, the contributions to the numerator and the denominator
look like
\begin{equation}
    \lim_{\tb\to\infty} \amuSUSY = \frac{\amured}{\Deltamured} \sim
    \frac{
      + \bigl( \chi^\pm\ \mathrm{term} \bigr) + \bigl( \chi^0\ \mathrm{terms} \bigr)}
    {
      - \bigl( \chi^\pm\ \mathrm{term} \bigr) + \bigl( \chi^0\ \mathrm{terms} \bigr)}
,
\end{equation}
i.e.\
both $\amured$ and $\Deltamured$ consist of
the chargino terms and the neutralino terms
whose diagrams are shown in figure~\ref{fig:diagrams}
in the mass insertion approximation.
The last fraction above is intended to mean that
the numerical values of the chargino terms in $\amured$ and $\Deltamured$
have opposite signs to each other
whereas the neutralino terms have the same sign.
When the five mass parameters are of similar sizes,
the chargino terms tend to dominate in both $\amured$ and $\Deltamured$
thereby resulting in negative $\amuSUSY$.
A way to turn around the sign would then be to let
the neutralino terms dominate or equivalently to
suppress the chargino terms in both $\amured$ and $\Deltamured$.
For this, one may suppress any of the propagators forming the loop
of the chargino graph in figure~\ref{fig:diagrams}(a),
as long as some of the neutralino diagrams survive.
\begin{figure}
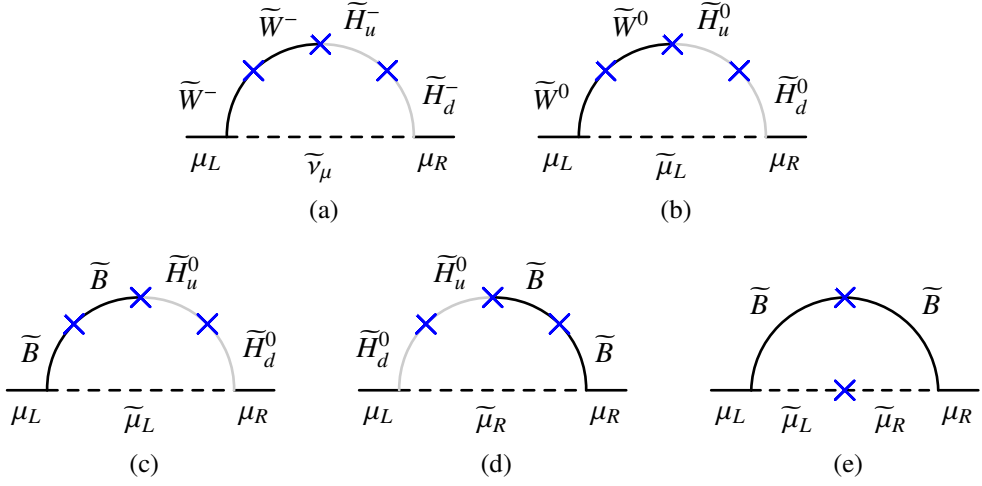

  \centering
  \graphsAsSubfigs{largemu}{(.8,,.8,,.8)}{black}{black}{black}{black}
  \caption{Same diagrams as in figure~\protect\ref{fig:diagrams}
    in the ``large-$\mu$ limit''.
    Grey lines represent mass-suppressed propagators.}
  \label{fig:diagrams large-mu}
\end{figure}
\begin{figure}
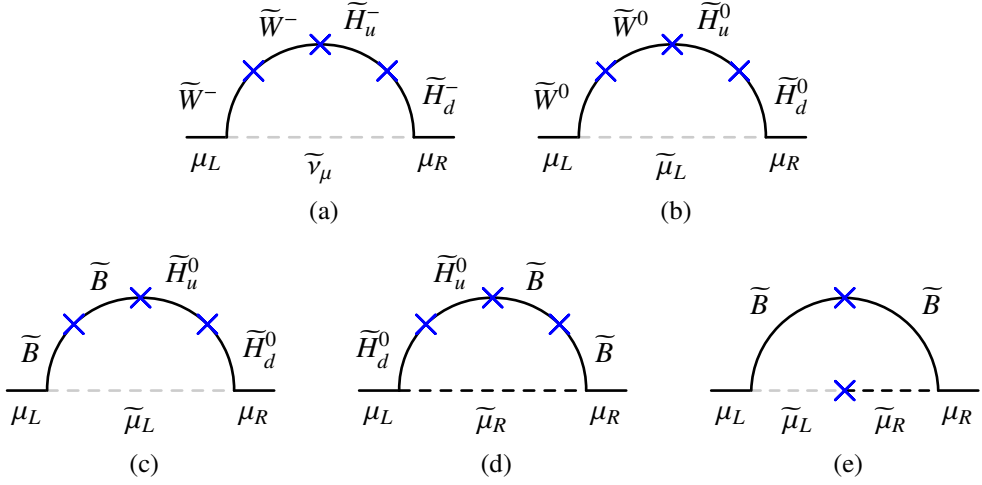

  \centering
  \graphsAsSubfigs{smuRdom}{black}{black}{black}{(.8,,.8,,.8)}{black}
  \caption{Same diagrams as in figure~\protect\ref{fig:diagrams}
    in the case of ``$\tilde{\mu}_R$-dominance''.
    Grey lines represent mass-suppressed propagators.}
  \label{fig:diagrams smuR-dominance}
\end{figure}
\begin{figure}
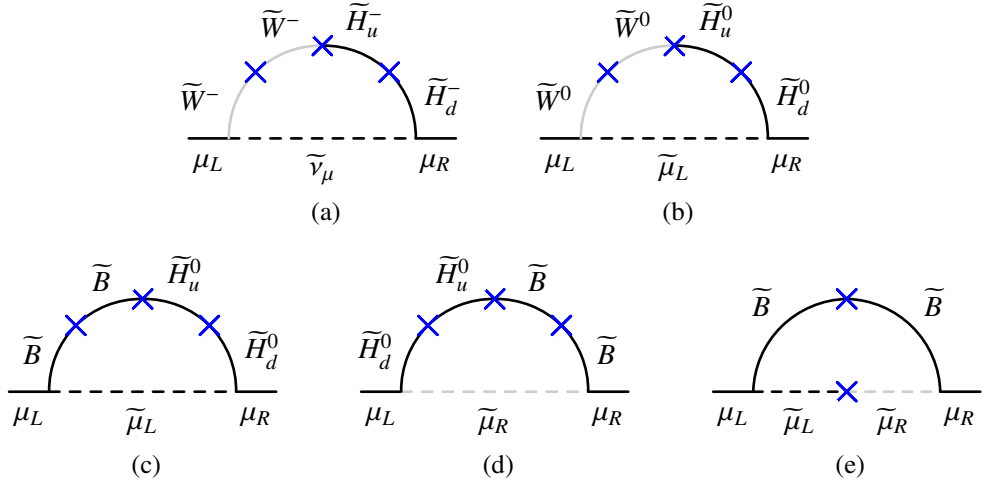

  \centering
  \graphsAsSubfigs{bhmLdom}{black}{black}{(.8,,.8,,.8)}{black}{(.8,,.8,,.8)}
  \caption{Same diagrams as in figure~\protect\ref{fig:diagrams}
    in the case where $\Mone, \mu, \ML \ll |\Mtwo|, \MR$.
    Grey lines represent mass-suppressed propagators.}
  \label{fig:diagrams etc}
\end{figure}
This leads to the following three solutions:
(a) ``large-$\mu$ limit'' which suppresses the Higgsino propagators
as shown in figure~\ref{fig:diagrams large-mu} while leaving
diagram~\ref{fig:diagrams large-mu}(e) unsuppressed,
(b) ``$\tilde{\mu}_R$-dominance'' which suppresses the sneutrino propagator
by raising $m_L$ as shown in figure~\ref{fig:diagrams smuR-dominance}
while maintaining the contribution from
figure~\ref{fig:diagrams smuR-dominance}(d),
(c) suppression of the wino propagators as in figure~\ref{fig:diagrams etc}
[as well as the $\tilde{\mu}_R$ propagators to isolate
diagram~\ref{fig:diagrams etc}(c)]
by raising $|\Mtwo|$ and $m_R$.
It turns out that solution (c) requires extreme mass hierarchies to work
in practice and is therefore less interesting than the other two options.

As explained above,
$\amuSUSY$ is invariant under a $\mu$ sign flip.
Furthermore, a simultaneous sign flip of both $\Mone$ and $\Mtwo$
leaves $\amuSUSY$ in~(\ref{eq:amuSUSYratio}) invariant,
as one can understand from
each diagram in figure~\ref{fig:diagrams} which picks up
the sign of either $\Mone$ or $\Mtwo$.
Taking advantage of these symmetries,
we shall assume in what follows that $\mu$ and $\Mone$ are positive
without loss of generality while leaving the $\Mtwo$ sign free.

\begin{figure}
  \centering
  \begin{subfigure}{0.5\textwidth}
  \includegraphics[width=\textwidth]{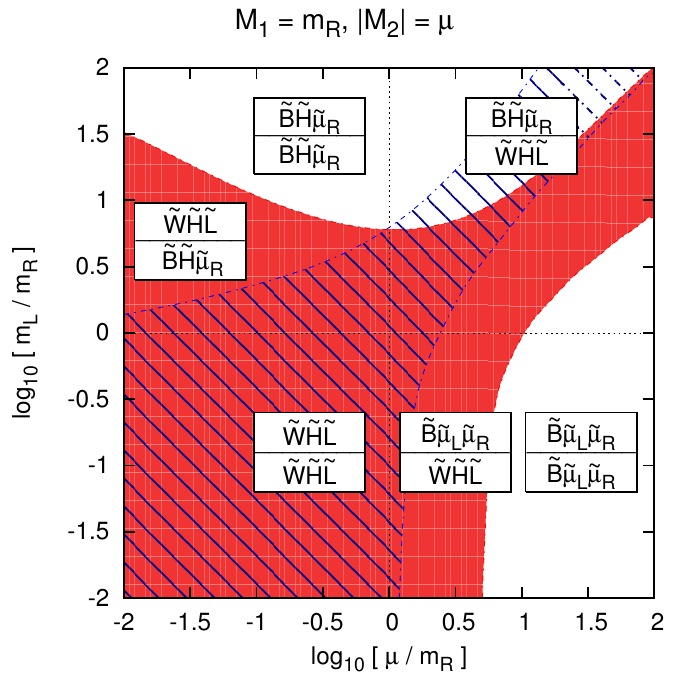}%
  \caption{}
  \end{subfigure}%
  \begin{subfigure}{0.5\textwidth}
  \includegraphics[width=\textwidth]{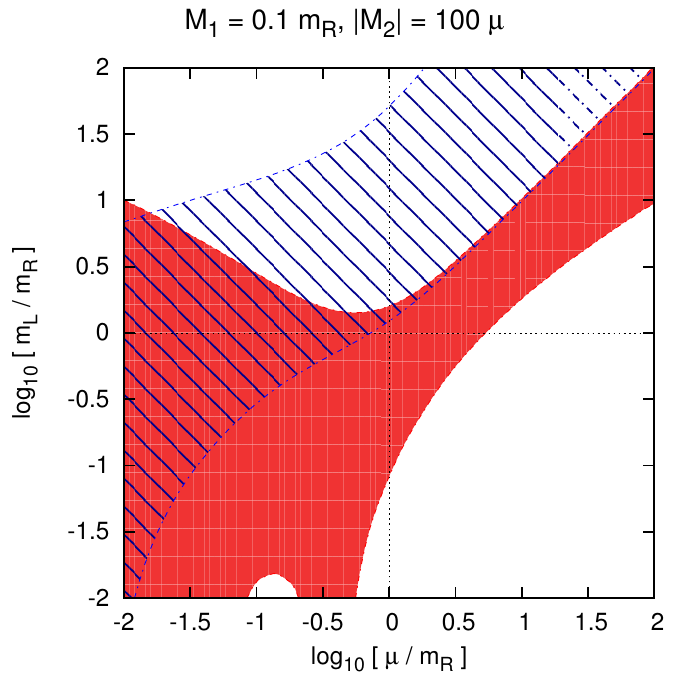}
  \caption{}
  \end{subfigure}
  \caption{Sign of $\amuSUSY$ on the plane of $(\MUE/\MR, \ML/\MR)$
    for the two signs of $\Mtwo$.
    The remaining two mass ratios are fixed as shown above each plot.
    The sign of $\amuSUSY$ in each region is: $+$ in white,
    $\sign(-\Mtwo)$ in red, $\sign(+\Mtwo)$ in hatched,
    $-$ in overlap.}
  \label{fig:signs}
\end{figure}

With the above qualitative observations in mind,
one can examine in more detail
how the sign of $\amuSUSY$ depends on the mass parameters.
Two plots are shown in figure~\ref{fig:signs}
employing the mass insertion approximation under which
the sign of $\amuSUSY$ is fully determined by the ratios
of the five mass parameters.
In figure~\ref{fig:signs}(a),
the origin at which the mass parameters are equal
leads to negative $\amuSUSY$ as already seen in~(\ref{eq:negativeamuSUSY}).
In the same plot,
one finds indeed the two types of regions where $\amuSUSY$ is positive
as expected from the aforementioned mass hierarchies:
the white regions on the right and around the upper border
correspond to the ``large-$\mu$ limit'' and ``$\tilde{\mu}_R$-dominance'',
respectively.
The less interesting solution~(c) is visible as a small white area
around the left part of the
bottom border of figure~\ref{fig:signs}(b), in which
$|\Mtwo|$ and $\MR$ are much larger than the other three mass parameters.

\begin{figure}[t]
  \centering
  \includegraphics[width=\textwidth]{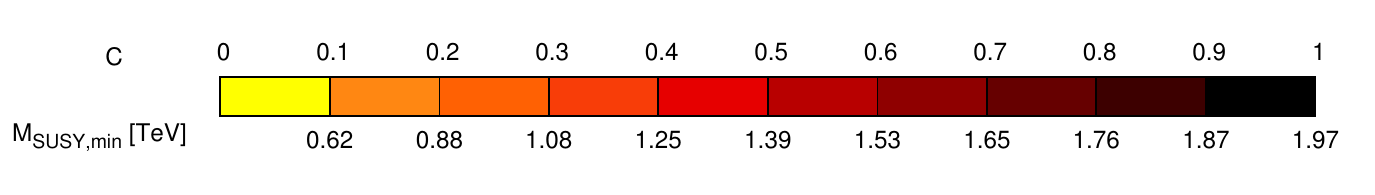}
  \\
  \begin{subfigure}{0.5\textwidth}
  \includegraphics[width=\textwidth]{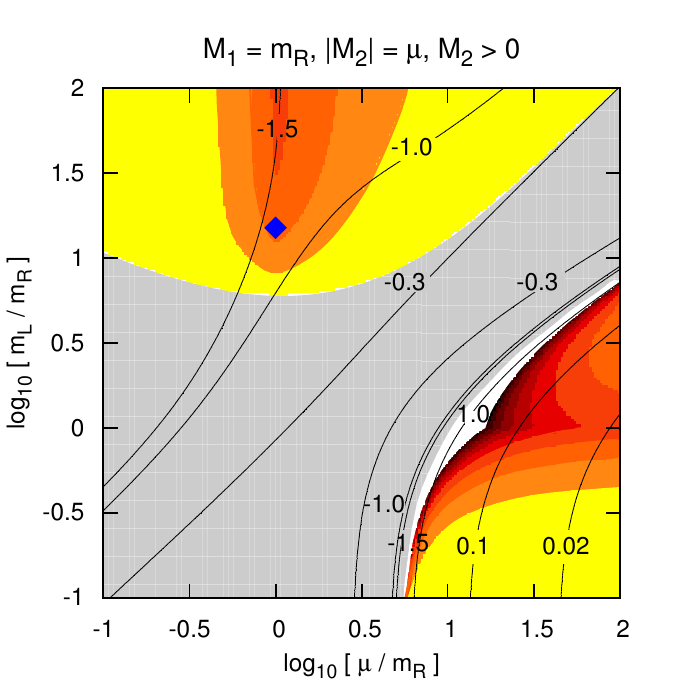}%
  \caption{}
  \end{subfigure}%
  \begin{subfigure}{0.5\textwidth}
  \includegraphics[width=\textwidth]{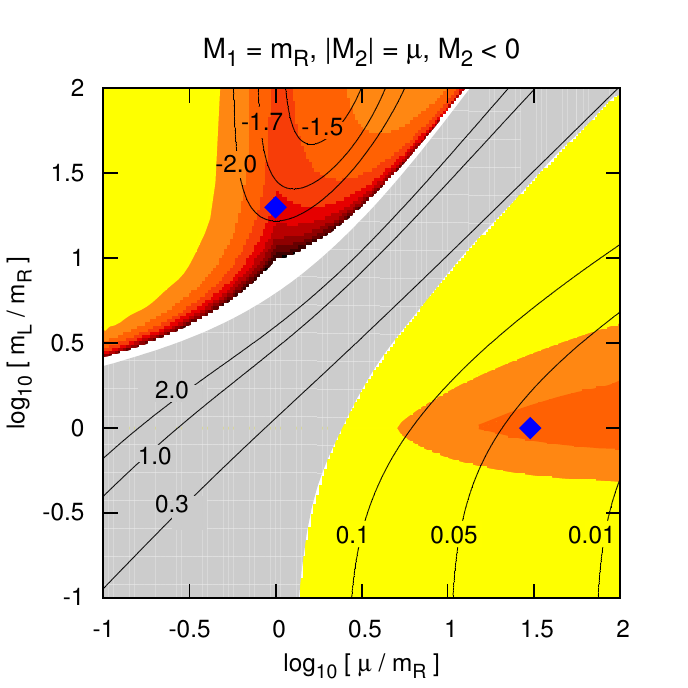}  
  \caption{}
  \end{subfigure}
  \caption{Values of $\MSUSYmin$ for the best fit of $\amuSUSY$
    are represented on the plane of $(\MUE/\MR, \ML/\MR)$
    as the gradation of colour,
    for (a) positive and (b) negative $\Mtwo$.
    The grey regions lead to negative $\amuSUSY$.
    The contour lines indicate $y_\mu$.
    In the white regions, $|y_\mu|$ is nonperturbatively large.
    The blue diamonds are the specimen points from
    table~\protect\ref{tab:specimen}.}
  \label{fig:gNY}
\end{figure}

Given a set of the mass parameter ratios leading to the desired sign of
$\amuSUSY$, its size can be adjusted to fit~(\ref{eq:amuexp-amusm})
by choosing an overall mass scale.
Obviously, the higher the mass scale is, the smaller $\amuSUSY$ becomes.
We parametrize the scale by $\MSUSYmin$,
the smallest of $\{ \mu, \Mone, |\Mtwo|, m_L, m_R \}$.
The results are presented in figure~\ref{fig:gNY}
in which one finds regions with colours corresponding to
$\MSUSYmin$ around $1\TeV$ or higher.
This reveals a promising possibility that supersymmetric particles
at the TeV-scale or higher can explain (\ref{eq:amuexp-amusm}),
which is the motivation to consider the infinite $\tb$ limit
in this work.

For this scenario to be viable,
it must meet all relevant constraints.
The lepton flavour violating process $\mu\rightarrow e\gamma$
can be suppressed
by assuming small enough slepton mixing between the first two generations.
The correlation between $\mu\rightarrow e\gamma$ and $a_\mu$
\cite{Kersten:2014xaa}
still holds for $\tb\to\infty$.
One can also satisfy $B$-physics constraints \cite{Altmannshofer:2010zt}.
The most dangerous decay mode is $B^+ \to \tau^+ \nu$ to which
the charged Higgs exchange contribution can be suppressed enough
by raising $M_{H^\pm}$ to a few TeV\@.
Pollution to $B_s \to \mu^+\mu^-$ and $B \to X_s\,\gamma$
can be suppressed with vanishing $A_t$ and
flavour-violating squark mass insertions.
For the former and the latter processes,
it further helps to raise the heavy Higgs and the squark masses,
respectively.
One can make
the lightest Higgs mass and decays SM-like
by staying in the decoupling regime \cite{Gunion:2002zf}.

\begin{figure}
  \centering
  \includegraphics[width=0.75\textwidth]{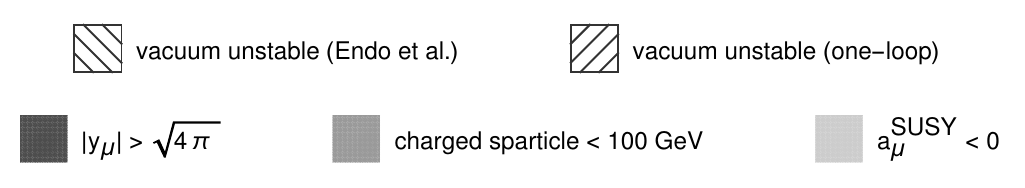}
  \\
  \begin{subfigure}{0.5\textwidth}
  \includegraphics[width=\textwidth]{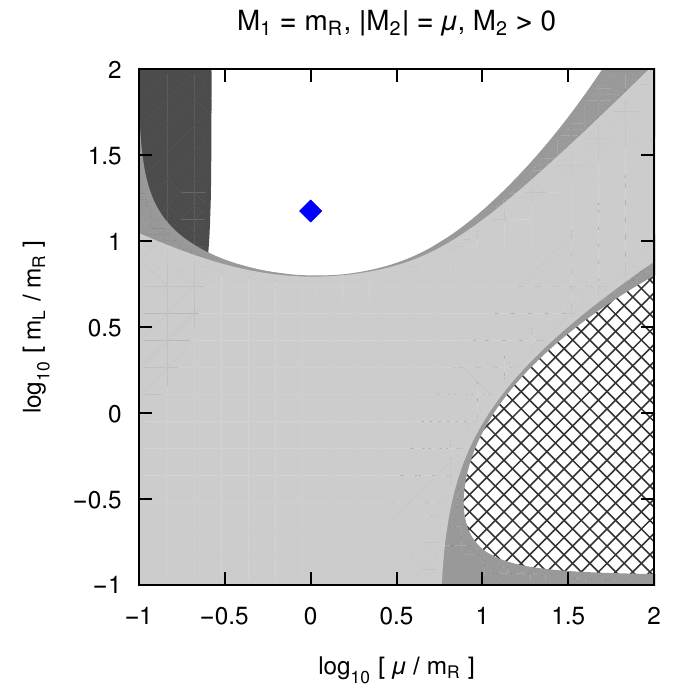}%
  \caption{}
  \end{subfigure}%
  \begin{subfigure}{0.5\textwidth}
  \includegraphics[width=\textwidth]{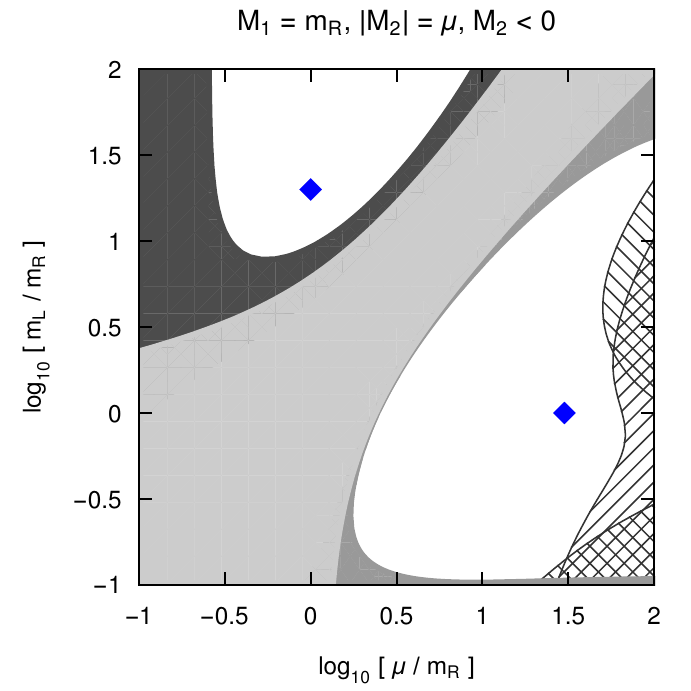}
  \caption{}
  \end{subfigure}%
  \caption{Regions on the plane of $(\MUE/\MR, \ML/\MR)$
    excluded by the constraints indicated in the legend,
    for (a) positive and (b) negative $\Mtwo$.
    The blue diamonds are the specimen points from
    table~\protect\ref{tab:specimen}.}
\label{fig:constraints}
\end{figure}

Moreover,
we impose the following constraints:
(a) charginos and smuons are heavier than $100\GeV$,
(b) $y_\mu$ is perturbative,
(c) our vacuum is stable or long-lived on the cosmological time-scale.
To evaluate the false vacuum lifetime,
we use the method from ref.~\cite{jaesmethod}.
Each of these requirements excludes some regions depicted in
figure~\ref{fig:constraints}.
The plots show that
great parts of the ``large-$\mu$'' region
for negative $\Mtwo$ as well as the ``$\tilde{\mu}_R$-dominance'' region
for either sign of $\Mtwo$ survive the constraints.
This means that there are indeed parameter volumes in which
TeV-scale supersymmetric particles can account for
the measured value of $a_\mu$.
We also check that $y_\tau$ and $y_b$,
the tau and the bottom Yukawas, can be perturbative
and consistent with the vacuum metastability.
Even though $m_\tau$, bigger than $m_\mu$,
might cause a worry about too large $|y_\tau|$,
there is room for perturbative radiative generation of $m_\tau$
since $(g_\tau - 2)$ does not need to be explained.
The even larger bottom quark pole mass can also be perturbatively generated
thanks to the gluino-loop contribution in addition to the other types of
loops shared by the tau self energy.

The concrete values of the five mass parameters at selected points
from figures~\ref{fig:gNY} and \ref{fig:constraints} are
listed in table~\ref{tab:specimen}.
\begin{table}
  \centering
\renewcommand{\arraystretch}{1.2}
\newcommand{\tabtev}{}
\begin{tabular}{ccccc|cc|c}
 \hline
 $\mu\tabtev$ & $\Mone\tabtev$ & $\Mtwo\tabtev$ &
 $\ML\tabtev$ & $\MR\tabtev$ & \rule[-.8em]{0pt}{2.3em} $\displaystyle{\amuSUSY}/{10^{-9}}$ & $\phantom{-}y_\mu$ &
 Characteristic\\ \hline
 $30$ & $1$ & $-30\phantom{-}$ & $1$ & $1$ & $2.80$ &
 $\phantom{-}0.04$ &large-$\mu$\\
 $15$ & $1$ & $-1\phantom{-}$ & $1$ & $1$ & $3.01$ & $\phantom{-}0.09$ &large-$\mu$ \\
 $1$ & $1$ & $\phantom{-}1\phantom{-}$ & $15$ & $1$ & $2.64$ & $-1.37$
 &$\tilde{\mu}_R$-dominance\\
 $1$ & $1$ & $\phantom{-}30\phantom{-}$ & $30$ & $1$ & $2.77$ &
 $-1.18$ &$\tilde{\mu}_R$-dominance\\
 $1.3$ & $1.3$ & $-1.3\phantom{-}$ & $26$ & $1.3$ & $2.90$ & $-1.89$
 &$\tilde{\mu}_R$-dominance\\
 \hline
\end{tabular}
  \caption{Specimen points leading to a reasonable fit of $a_\mu$.
  Masses are in TeV\@.}
  \label{tab:specimen}
\end{table}
The first two points belong to the ``large-$\mu$ limit''
with negative $\Mtwo$,
and the rest to the ``$\tilde{\mu}_R$-dominance'' regime
with either sign of $\Mtwo$.

\begin{figure}
  \centering
  \begin{subfigure}{0.5\textwidth}
    \includegraphics[width=\textwidth]{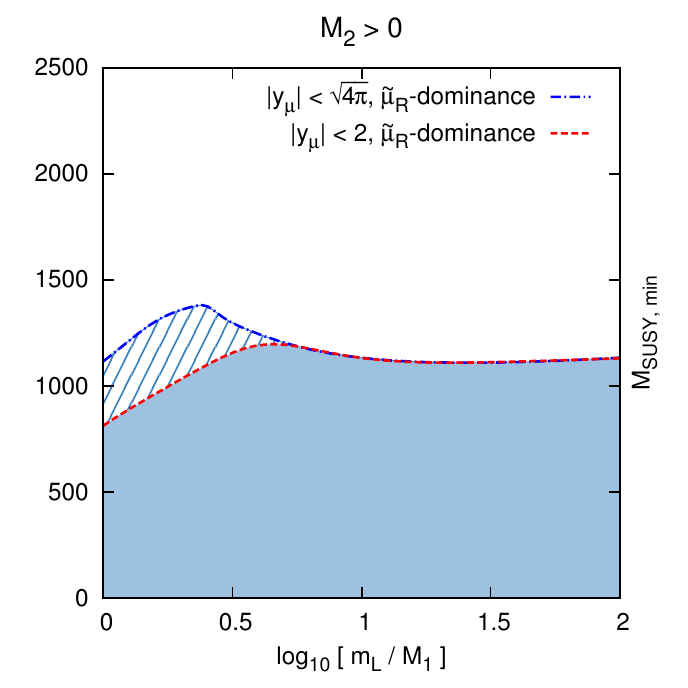}%
  \caption{}
  \end{subfigure}%
  \begin{subfigure}{0.5\textwidth}
    \includegraphics[width=\textwidth]{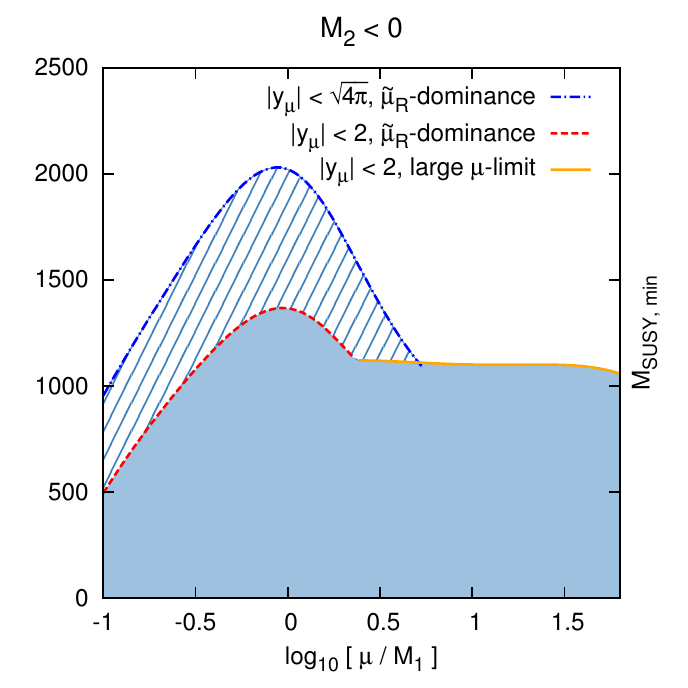}
  \caption{}
  \end{subfigure}%
  \caption{Maximum $\MSUSYmin$ that can fit $a_\mu$ as a function of
  (a) $\ML/\Mone$ for $\Mtwo > 0$ and (b) $\mu/\Mone$ for $\Mtwo < 0$.
  Each line style indicates to which regime the point belongs as well as
  the cutoff on $|y_\mu|$.
  The vacuum metastability is required.}
  \label{fig:Scanplots}
\end{figure}

To find maximal $\MSUSYmin$ which can fit $a_\mu$,
we explore the full five-dimensional parameter space,
relaxing the equalities, $\Mone = \MR$ and $|\Mtwo| = \mu$,
which we assumed for the sake of
planar presentation of the preceding plots.
The resulting maximum $\MSUSYmin$ is plotted in figure~\ref{fig:Scanplots}.
The plots reveal that
the relaxation of the above mass equalities allows
$\MSUSYmin$ to be slightly higher than before.

For a compact summary, we derive a formula
resembling~(\ref{eq:negativeamuSUSY}),
\begin{equation}
  \amuSUSY \approx 37 \times 10^{-10} \,
  \left(\frac{\unit{1}{\tera\electronvolt}}{\MSUSY}\right)^2 ,
\end{equation}
which applies if either
$|\mu|\gg |\Mone|=m_L=m_R\equiv M_{\text{SUSY}}$ or
$m_L\gg |\mu|=|\Mone|=m_R\equiv M_{\text{SUSY}}$.
These are two mass hierarchies representative of
the ``large-$\mu$ limit'' and the ``$\tilde{\mu}_R$-dominance'',
respectively.


In summary,
we considered the infinite $\tb$ limit of the MSSM
as a possibility to account for the discrepancy
between the experimental value and the SM prediction
of $a_\mu$
even if the supersymmetric particles are as heavy as $1 \TeV$ or higher.
The motivation was the observation that fully radiative muon mass generation
would imply large new physics effects on $a_\mu$.
We found two successful types of mass hierarchies, the ``large-$\mu$ limit'',
and the ``$\tilde{\mu}_R$-dominance'',
which allowed us to achieve the goal.
We took into account phenomenological constraints from
collider searches, flavour and Higgs physics, as well as
theoretical constraints from perturbativity and vacuum metastability.
For more details of the analysis,
we refer the reader to ref.~\cite{Bach:2015doa}.

For those $a_\mu$ enthusiasts,
we add a note on
GM2Calc, a calculator of the MSSM contributions to $a_\mu$ \cite{gm2calc}.
It can approximate the infinite $\tb$ limit
as it works for arbitrarily high $\tb$.


J.P. acknowledges support from the MEC and FEDER (EC) Grant
FPA2011--23596 and the Generalitat Valenciana under Grant PROMETEOII/2013/017.

\end{document}